\documentclass[doublecol]{epl2}
\usepackage{graphicx}

\title{The origin of the strain-rate discontinuity in 2D foam rheometry
with circular geometry}
\author{Denis Weaire, Robert J. Clancy and Stefan Hutzler\footnote{email address of communicating author: stefan.hutzler@tcd.ie}}
\shortauthor{Weaire \etal}

\institute{
   School of Physics, Trinity College Dublin, Dublin 2, Ireland\\
}

\pacs{82.70.Rr}{Foams}
\pacs{83.80.Iz}{Foams, rheology}
\pacs{83.60.La}{Yield stress (rheology)}

\abstract{
The observed discontinuity in strain-rate for a two-dimensional foam undergoing
shear in a (circular) Couette system is explained in terms of the
continuum (Herschel-Bulkley) model. It is attributed to the finite difference
between yield and limit stress.
}

\begin{document}

\maketitle

\section{Introduction}

\begin{figure}[tbp]
\begin{center}
\includegraphics[width = 8cm]{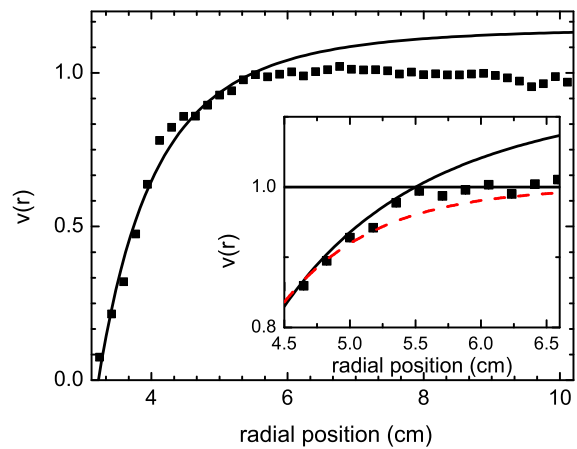}
\end{center}
\caption{Reprinted figure 5 from \cite{Dennin08}. 
In the case of circular (Couette) geometry, the scaled azimuthal
steady-state velocity
$v(r)/(r \Omega)$, where $r$ is the radial position and $ \Omega$ is the rotation rate of the outer cylinder,
may become constant at some internal point (see also figure
\ref{f:geometry}). 
In the data of
\cite{Lauridsen04,GilbrethEtal06,Dennin08},
taken for a moving outer boundary and a fixed inner one,
there is a discontinuity of $d(v(r)/(r \Omega))/dr$ at this point. {\em
Note that this published figure uses the symbol $v(r)$ to denote $v(r)/(r
\Omega)$ in the notation of the present paper.} The solid and dashed
lines show least-square fits to power-law and exponential, respectively. 
This figure is used
with permission from M. Dennin \cite{Dennin08}, 
Copyright 2008 by IOP Publishing Ltd.}
\label{f:dennin_data}
\end{figure}

Many of the existing observations in 2D foam rheology 
\cite{Debregeas01,Lauridsen02,Lauridsen04,Wang06} may be
accounted for within the general continuum model that incorporates bulk
dissipation and, where appropriate, wall drag 
\cite{Janiaud06,ClancyEtal06,JaniaudEtal07,WeaireEtal08b}. One feature that has 
so far escaped such an
explanation is the discontinuity of the derivative of velocity (or of
the shear rate), reported by the group of 
M. Dennin \cite{Lauridsen04,GilbrethEtal06} for the case of a 2D foam in a 
circular (Couette) rheometer.  Figure \ref{f:dennin_data} shows the data.
These measurements were made using the variety of 2D foam that is also
known as the Bragg raft.

\begin{figure}[tbp]
\begin{center}
\includegraphics[width =
8cm]{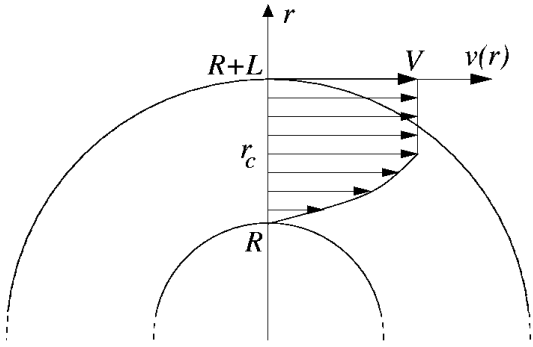}
\end{center}
\caption{Sketch of velocity profile for a moving outer boundary in the case
where continuous shear is confined to $R<r<r_c$.}
\label{f:geometry}
\end{figure}
                                                                                
This derivative in question 
is everywhere continuous in the existing continuum model.  
In the present paper we show that a natural refinement of that model gives
rise to just what is observed, and our simulations of the effect offer
opportunities for further experimental tests.

The refinement that is required is the incorporation of a stress-strain
relation that has a peak or overshoot before it settles down to its
asymptotic value. That is, there is a {\em yield} stress that is in excess of
the eventual {\em limit} stress, and it is the difference between 
these quantities
that relates to the discontinuity (see
Fig.~\ref{f:yielding_function}). This has for long been a familiar feature
of measurements and simulations of foam rheology (e.g.\cite{KhanEtal88}).

However, as elsewhere in the physics of foams, the outcome depends on the
detailed history (and even prehistory) of the experiment, so it is
essential to study simulations of the velocity profile in various cases.  
We shall see that this discontinuity does indeed depend on the
experimental protocol.

The effect should not, of itself, be restricted to the 2D foam system in
question, but we are not aware of any reference to it in 3D foams. In any
case, it offers further support for the continuum model in the present
context.

In most of what we present, the original primitive version of the
continuum model is used \cite{Janiaud06}. In this, linear forms are used 
for bulk dissipation and wall drag terms. Later we will indicate the 
straightforward modification of the main result that arises from the 
introduction of nonlinear forms, corresponding
to the later and more realistic development of the model
\cite{WeaireEtal08b}.

\section{The continuum model with distinct yield and limit stresses}

Clancy {\em et al.} \cite{ClancyEtal06}  have provided an extensive analysis of the 2D foam
continuum model for circular (Couette) geometry. It is a much richer
scenario than that of simple shear between straight boundaries, although
the two are closely similar in some limits. Localisation due to wall drag
is combined with geometric effects in this case, so that there are several
qualitatively distinct forms of the velocity profile, according to the
dimensions of the system, the parameters of the model, the imposed
velocity, and whether it is applied to the inner or outer boundary.

A particular feature of the circular case is the possibility of the existence
of a range of radial positions 
, $r$,
 at which the strain rate is zero (implying
rigid body rotation), as in Figure \ref{f:dennin_data} and the sketch in Figure \ref{f:geometry}.

The present topic relates to just this feature, specifically to the 
boundary point $r=r_c$ between the ranges of zero and non-zero shear rate.  

We follow closely the formalism and algorithms of the previous 
work \cite{Janiaud06,ClancyEtal06,JaniaudEtal07,WeaireEtal08b}. 
In the analysis presented here we will restrict ourselves to a moving outer
cylinder to match the experimental conditions of Figure
\ref{f:dennin_data}.
The local strain rate $\dot{\gamma}(r)$ will thus be positive throughout.

We begin with an elementary
steady-state
 treatment that yields a formula for the 
discontinuity, before embarking on a time-dependent simulation.

\begin{figure}[tbp]
\begin{center}
\includegraphics[width = 8cm]{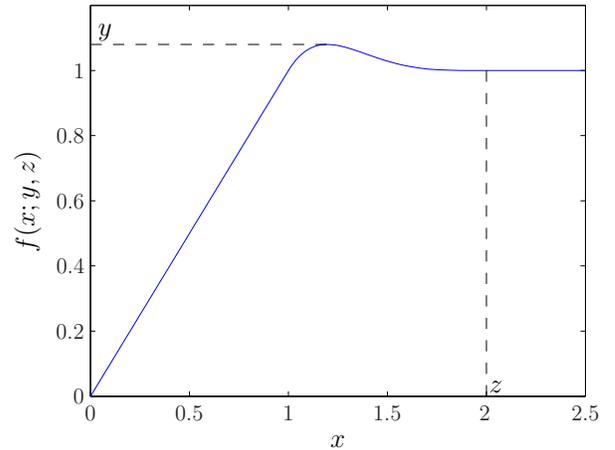}
\end{center}
\caption{Plot of the stress-strain function $f(x;y,z)$
of eqn.~\ref{e:yielding_function} where we have set
$y=\frac{\sigma_Y}{\sigma_L}=1.08$ and $z =  \frac{\gamma_L}{\gamma_Y}= 2$
for illustrative purposes. Numerical evaluation of eqn.~\ref{e:beta} then results in
$\beta  = 4.1076$.
The form of $f(x;y,z)$ as given by eqn.~\ref{e:yielding_function}
is only valid for
the case where the strain rate is always positive.}
\label{f:yielding_function}
\end{figure}

The constitutive relation expressing local stress $\sigma(r)$ as a function
of radial position $r$, local strain
$\gamma(r)$ and local strain rate $\dot{\gamma}(r)$ is given by
\cite{Janiaud06,ClancyEtal06,JaniaudEtal07,WeaireEtal08b}
\begin{equation}
\sigma=\sigma_Lf(\gamma/\gamma_Y;\sigma_Y/\sigma_L,\gamma_L/\gamma_Y)+\eta \dot{\gamma}
\label{e:constitutive}
\end{equation}
where we call the first term on the right-hand side the {\em
elastic-plastic stress}.
Yield and limit stress are denoted by
$\sigma_Y$ and $\sigma_L$, respectively,
$\gamma_L$ and $\gamma_Y$ are yield and limit strain. The 
viscosity component of stress (also called
{\em consistency})   is denoted by
$\eta$. 
As a function $f$ representing the scaled quasi-static stress-strain
relation we have arbitrarily chosen
\begin{equation}
f(x;y,z)= \left\{ \begin{array}{ll}
x & \textrm{for $x<1$} \\
1+(x-1)\left(\frac{z-x}{z-1}\right)^\beta & \textrm{for $ 1<x<z$} \\
1 & \textrm{for $x>z$} 
\end{array} \right.
\label{e:yielding_function}
\end{equation} 
where $\beta$ is fixed in terms of $y$ and $z$ by
the value of the maximum of $f$ (w.r.t. $x$) for $1<x<z$,
\begin{equation}
\frac{\beta^\beta}{(1+\beta)^{1+\beta}}=\frac{y-1}{z-1}
\label{e:beta}
\end{equation}
See Fig.~\ref{f:yielding_function} for a plot of the yielding curve used in our simulations in our time-dependent simulations, described below. 

As the strain-rate discontinuity is reported for bubble rafts
\cite{Lauridsen04,GilbrethEtal06}, the
expression for the wall drag in the continuum model \cite{Janiaud06} is omitted. However,
the key result of this paper would not be affected by it. 

Following
from appendix A.3 of \cite{ClancyEtal06} we thus have (in the absence of
inertia) 
\begin{equation}
\frac{1}{r^2}\frac{\partial}{\partial r}r^2\sigma=0
\end{equation}
and so the radial variation of stress is given by
\begin{equation}
\sigma=\frac{A}{r^2}
\label{e:radial_stress}
\end{equation}
where $A$ is a constant, fixed, for example,  by the stress on the moving boundary.


The matching conditions for the steady state profile at the boundary point 
$r_c$ are as follows.

\begin{itemize}
\item Velocity $v(r)$ is continuous (otherwise the strain rate would be
infinite).
\item Total stress $\sigma$ is continuous since it is given by
eqn.~\ref{e:radial_stress}.
\end{itemize}

The second condition is most relevant here. On the side of $r_c$ that is 
undergoing continuous shear with rate $\dot{\gamma}(r)$, Equation
(\ref{e:constitutive}) reduces to:

\begin{equation}
\sigma=\sigma_L+\eta \dot{\gamma},
\label{e:inner}
\end{equation}
where $\sigma_L$  is the limit stress (Fig. \ref{f:yielding_function}). 

On the other 
side of $r_c$, 
there is no continuous shear, so the total stress is given by the
elastic-plastic stress (first
term in Equation \ref{e:constitutive}). {\em Prima facie}, it appears that
 the value
of the elastic-plastic stress
 at $r_c$ can be anywhere in the range $0 < \sigma < \sigma_Y$, and we here
assume the value $\sigma = \sigma_Y$ is appropriate. The next
section describes time-dependent simulations that validate
this choice for the usual experimental procedure.

Equating the total stress on the two sides of $r_c$, we obtain

\begin{equation}
\dot{\gamma}(r_c)=(\sigma_Y-\sigma_L)/\eta.
\label{e:stress_rc}
\end{equation}

Since the yield stress exceeds the limit stress, $\sigma_Y > \sigma_L$, the strain-rate
on the shearing side
is finite at $r_c$, leading to a discontinuity in the strain-rate at that point,
hence to a discontinuity in the azimuthal velocity gradient 
at this point.

\section{Range of possible magnitudes of the discontinuity}
The argument above allows {\em any} value of the elastic-plastic stress at
$r_c$ that is less than $\sigma_Y$, however, the possibility of having
$\sigma < \sigma_L$ may be dismissed as follows.

Such a value would imply, through Equation \ref{e:inner}, a {\em
negative} value 
of $\dot{\gamma}$ for values of radii less than, but close to $r_c$. 
Assuming 
continuity of
 the velocity, this in term implies a
velocity maximum in this range, at some value $r_{max}$. 
There is no force on an
element at that radius due to the elastic-plastic stress (in the shearing region the elastic-plastic stress must be the limit stress
everywhere, because we are dealing with the steady state),
but there is a force present due to $\frac{\partial^2 v(r)}{\partial r^2}$. 
Since this cannot be balanced, a value of $\sigma$ which is less than $\sigma_L$
is thus not possible.

\section{Time-dependent simulations}
\label{s:time_dependency}

To explore this further, we resort to the full implementation of the 
time-dependent model \cite{ClancyEtal06}, and simulate an experiment in which the system 
starts from static stress-free equilibrium, and the outer boundary is set 
in motion with a fixed velocity.

\begin{figure}[tbp]
\begin{center}
\includegraphics[width =
8.5cm]{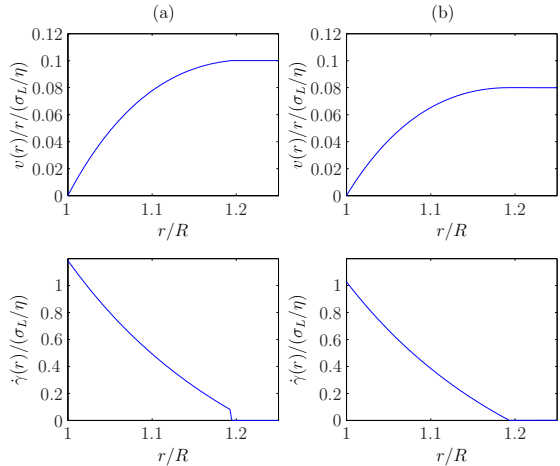}
\end{center}
\caption{Exemplary calculations for the continuum model in the circular
geometry of figure \ref{f:geometry}, with $L=1.25 R$ and the stress-strain
function of figure ~\ref{f:yielding_function}. (a) shows the steady state for both local $v(r)/r$ and
strain-rate $\dot{\gamma}$ (both expressed in dimensionless units) 
for an outer cylinder
rotating at a rate $\Omega_1=\frac{v(R+L)}{R+L} = 0.1$. The
position $r_c/R\simeq 1.19$, where the
strain-rate has a discontinuity separates the shearing ($r<r_c$) from the
non-shearing region ($r>r_c$). A decrease of the rotation rate of the outer
cylinder to $\Omega_2=0.08$ eventually leads
to the steady-state profiles shown in (b). Note that the discontinuity in the
strain-rate has now vanished.
}
\label{f:movies}
\end{figure}

Figure \ref{f:movies} (a) shows angular
velocity and strain-rate once a steady state is reached \footnote{Movies of the simulations can be downloaded at
http://www.tcd.ie/physics/foams/publications.php}. The
strain-rate is discontinuous at $r_c$. Also, in agreement with our
argument above and Eqn.~\ref{e:stress_rc}, it remains positive in the
vicinity of $r_c$, and is given by $\dot \gamma(r_c) / (\sigma_L \eta) =
0.08$ for the choice of $\sigma_Y = 1.08 \sigma_L$ in the simulations shown.
Upon decrease of the
velocity of the outer boundary  a new steady
state is reached with velocity and strain-rate profiles as shown in Figure
\ref{f:movies} (b).  Our simulations show
the absence of the discontinuity of the strain rate for this scenario.

As in previous work \cite{ClancyEtal06} our algorithm for the
implementation of the continuum model is not stable unless allowance is
made for the possibility of negative strain rate $\dot{\gamma}$, even
though this would appear not to occur in the solution represented by the
numerical results. We still do not have any explanation of this
technicality.

\section{General continuum model}

In the generalisation of the continuum model we introduce a non-linear
dependency of the strain rate dependent term in the constitutive relation
of Equation
(\ref{e:constitutive}),
\begin{equation}
\sigma=\sigma_Lf(\gamma/\gamma_Y;\sigma_Y/\sigma_L,\gamma_L/\gamma_Y)+\eta'
\dot{\gamma}^a,
\label{e:constitutive_nonlin}
\end{equation}
where the exponent $a$ is called the
Herschel-Bulkely exponent \cite{WeaireEtal08b}.
Using the same argument as for the linear case, we again find a
discontinuity in the strain rate at the boundary point $r_c$ between the
ranges of zero and non-zero strain-rate. The value of strain rate at this point
is given by 
\begin{equation}
\dot{\gamma}(r_c)=\left(\frac{\sigma_Y-\sigma_L}{\eta'}\right)^{1/a}.
\label{e:stress_rc_nonlin}
\end{equation}

\section{Conclusion}

The overshoot that plays a central role in the above arguments is quite 
familiar in foam physics, but is nevertheless not understood in any 
detailed sense. It is attributable to the change of structure under 
imposed continuous shear \cite{WeaireEtal92}.

The meaning of this is not entirely self-evident. Suppose we shear a foam 
considerably, and then reduce stress to zero. Of course, the new structure 
is microscopically different from the original one, but it is also 
different in an average, statistical sense. It will, for example, be 
anisotropic, even if  the original sample was not. 

The explanation of the shear rate discontinuity, as presented here, may not
be unique but seems compelling. It
invites further tests, by varying the experimental parameters and, in 
particular, following the protocol which is predicted to eliminate the 
discontinuity.

This particular effect is not confined to two-dimensional samples.
Accordingly, it may well have been noted 
in 3d rheometry at some earlier stages, but we are unaware of any direct such
observations. In the present context, its explanation lends 
further support to the continuum model as the natural ``mean field'' 
model to be applied to  shear localisation in 2D foams, at least as a first
approximation.
It also offers a test for further experiments, to verify that the
discontinuity may be eliminated by a different protocol, as seen in our
time-dependent simulations.

\section{Acknowledgments}

We are grateful to M Dennin for discussions an permission to reproduce 
Figure \ref{f:dennin_data}.
This research was supported by the European Space Agency (MAP
AO-99-108:C14914/02/NL/SH and (MAP
AO-99-075:C14308/00/NL/SH)
and the  COST P21 action (European Science Foundation)
on ``The
physics of droplets''.

\end{document}